# High-temporal-resolution electron microscopy for imaging ultrafast electron dynamics

M. Th. Hassan*, J. S. Baskin*, B. Liao, and A. H. Zewail†

Physical Biology Center for Ultrafast Science and Technology, Arthur Amos Noyes Laboratory of Chemical Physics, California Institute of Technology, Pasadena, CA 91125, USA

† Deceased, August 2016.

Ultrafast Electron Microscopy (UEM) has been demonstrated to be an effective table-top technique for imaging the temporally-evolving dynamics of matter with subparticle spatial resolution on the time scale of atomic motion. However, imaging the faster motion of electron dynamics in real time has remained beyond reach. Here, we demonstrate more than an order of magnitude (16 times) enhancement in the typical temporal resolution of UEM by generating isolated ~ 30 fs electron pulses, accelerated at 200 keV, via the optical-gating approach, with sufficient intensity for efficiently probing the electronic dynamics of matter. Moreover, we investigate the feasibility of attosecond optical gating to generate isolated subfemtosecond electron pulses, attaining the desired temporal resolution in electron microscopy for establishing the "Attomicroscopy" to allow the imaging of electron motion in the act.

Ultrafast Electron Microscopy (UEM) and Diffraction (UED) are pivotal tools allowing for real-time imaging of chemical reactions,[1-3] surface dynamics,[4,5] and structural phase transformations[6-8]. Moreover, the recent advances allows the generation of ultrafast mega-electron-volt (MeV) electron pulses to study the molecular structural changes in gas phase[9,10]. However, the imaging and visualization of electronic motion, which provides insight into electron dynamics in atoms[11,12], molecules,[13-15] and condensed matter systems,[16] has not hitherto been achieved due to the lack of the requisite high temporal resolution.

The temporal resolution in ultrafast electron microscopy depends on the duration of the electron pulse which is finite due to the initial energy dispersion and the space-charge effect. Therefore, many electron pulse compression techniques have been developed for generating short electron pulses. Recently, sub-100 fs electron pulses have been obtained utilizing radio frequency (RF), microwave, and terahertz field compression techniques[17-19]. However, in ultrafast dynamics measurements that have been carried out so far, the electron pulse durations typically ranges from a picosecond to several hundreds of femtoseconds,[1,3,8,20,21] which is insufficient to resolve electron dynamics of matter. Hence, the generation of electron pulses with a few tens of femtosecond to hundreds of attosecond duration is necessary for freezing and imaging the electron motion in real-time, as theoretically demonstrated in atoms and molecules[22], and condensed matter[23].

**Optical gating of ultrafast electron pulses**

In this work we demonstrate a major enhancement in the temporal resolution in UEM by generating isolated ultrashort (30 fs) electron pulses utilizing the optical gating methodology. This approach is based on the Photon Induced Near-field Electron Microscopy (PINEM)[24] in which the optical photon-electron coupling takes place in the presence of nanostructures when the energy-momentum conservation condition is satisfied[24-26]. Recently, this coupling has been exploited for coherent control of the quantum state of free-electron populations in an electron beam[27]. Also, it has been used for controlling and streaking the free electron beam by sub-cycle laser field[28]. In this coupling process, the inelastic interaction between the electron pulse and the laser photons leads to gain/loss of photon quanta by some of the electrons in the electron packets, which can be resolved in the electron energy spectrum consisting of discrete peaks, spectrally separated by multiples of photon energy (nℏω), on both sides of the zero loss peak (ZLP). Since, the electrons can gain or lose energy due to the coupling only in the presence of photons, the optical laser pulse acts as a "temporal gate" of these electrons. These gated electrons can be filtered out, providing a dramatic enhancement of the temporal resolution in electron microscopy for exploring ultrafast dynamics of matter triggered by another ultrashort optical laser pulse in different UEM modes (i.e. diffraction, electron spectroscopy and/or direct imaging)[24-30].

The striking features of this approach are; first, the temporal gating window and duration of the generated electron pulse is limited only by the optical "gating" pulse, when utilizing a nanostructure coupling system with intrinsic dynamics life-time shorter than the gating window, which in principle can be on the order of subfemtosecond time scale, utilizing the optical attosecond pulse[11], second, the phase drift between the optical "gating" pulse (and consequently gated-electron pulse) and the laser pulse triggering dynamics can be also on the subfemtosecond time scale, thanks to the optical phase locking technique[11], third, the "gated" electron pulses have enough intensity to practically image sample dynamics since the "original" electron pulse is not restricted to be in the single electron-regime because the "gated" electron pulse is generated after acceleration and at the position of the sample under study. Therefore, this approach is a promising technique to break the limits of the temporal resolution not only in the ultrafast electron microscopy, but also for the free-electron lasers[28,31], where extreme time-resolution is difficult to achieve.

In contrast, electron pulse compression by radio frequency (RF) and microwave utilized to generate sub-100 fs electron pulses have the limitation of the pulse intensity, phase drift and timing stability[32,33]. Although, in the recent reported terahertz field compression technique,[19] the phase drift stability was enhanced to achieve a ~4 fs jittering stability, the demonstrated ~75 fs electron pulse remains insufficient to record snapshots of the fast electron motion which last only hundreds of attosecond to a few tens of femtosecond. Other proposed approaches to achieve attosecond resolution[27,28,34] anticipate obtaining a train of short electron pulses which cannot be utilized in practical time resolved electron microscopy measurements. For these reasons, the optical-gating approach is the choice for generating isolated electron pulses with extreme limits from a few tens of femtosecond to attosecond time scale and attaining the temporal resolution of electron motion in electron microscopy, opening the door for vast femtosecond and attosecond stroboscopic imaging applications.

**Generation and temporal characterization of isolated 30 fs gated electron pulse**

In our experiment (setup is shown in Fig. 1, more details in Method), the visible pulse ("gating" pulse) enters the microscope, and together with the ultrafast "original" electron pulse (generated by photoemission), illuminates the nanostructure specimen (gold nanoparticle). The "gating" pulse is kept at low power (~1.8 mJ/cm$^2$) to avoid saturation. At the spatiotemporal overlapping of the two pulses, the coupling between the pulses takes place. The signature of this coupling can be revealed by measuring the electron energy spectrum using the electron energy spectrometer attached to the microscope. The measured electron spectrum is shown by the shaded blue curve in Fig. 2a. The shaded white reddish curve represents the ZLP spectrum in the absence of coupling. The side peaks in the shaded blue curve constitute the "gated" electron spectrum.

The temporal profile of the "original" electron pulse can be characterized by means of cross-correlation between the "gating" visible pulse and the electron pulse; since the "gating" pulse duration is much shorter, the cross-correlation reflects directly the electron pulse temporal profile. This use of an ultrashort "gating" optical pulse introduces a simple technique in electron pulse metrology. To this end, the electron energy spectra are measured as a function of the delay time between the electron and the "gating" pulse ($\tau_{Vis}$). The recorded spectrogram is depicted in Fig. 2b. The cross-correlation temporal profile, retrieved by integrating the gated electron spectra (side peaks) at each delay time($\tau_{Vis}$), and its fitting (the open black circles and red curve, respectively) are shown in Fig. 2c. The pulse duration of these "original" electron pulses is on the order of 500 fs.

For generating the isolated ultrashort "gated" electron pulse with maximum counts, the "gating" pulse is kept at the optimum temporal overlap ($\tau_{Vis} = 0$ fs) with the few hundred femtosecond electron pulse. The pulse duration of these gated electrons is comparable to the temporal profile of the "gating" optical pulse (~30 fs, characterized by the autocorrelator), which is shown by the white dotted line and white shading in Fig. 2c. The temporal profile of these gated electrons is retrieved by the same principle of the electron-photon coupling cross-correlation utilizing a second laser pulse.

The principle of this measurement is illustrated in Fig. 3a. The time delay of the second laser pulse (here, $\tau_{NIR}$ of the NIR pulse (~1.4 mJ/cm2)) is scanned across the "gated" and "original" electron pulses and can couple to both of them as a function of temporal overlap. The temporal profile of "gated" electron pulse can be retrieved from the coupling cross-correlation temporal profile between the "gated" electron and the NIR pulse given the fact the pulse duration of the NIR pulse is known. Fig. 3b shows the electron energy spectra of: (i) the "original" few hundred femtosecond electron pulse (ZLP) (orange shaded curve) (ii) the coupling between the "gating" visible pulse and the "original" electron pulse (black curve), and (iii) the coupling between the NIR laser pulse and both the "original" and "gated" electron pulses at $\tau_{NIR} = 0$ fs (red curve). The last coupling energy spectrum is measured as a function of $\tau_{NIR}$ and the acquired spectrogram (with ZLP suppressed) is plotted in Fig. 3c. In fact, the information about the "gated" electron pulse duration is carried on the spectral modulation of this spectrogram, which can be retrieved by tracing the modulation of the ZLP spectrum. This modulation reflects the change in the intensity of the entire coupling spectrum on both sides of the ZLP. The inverse of the ZLP intensity change is plotted in connected open black circles in Fig. 3d, which can be obtained by integrating the ZLP areas as a function of the delay time $\tau_{NIR}$ (average of 6 scans) from the recorded spectrogram. This curve carries the information of the cross-correlation coupling between the "original" electron and NIR pulse together with the cross-correlation between the "gated" electron and NIR laser pulse. Then, to extract the "gated" electron temporal profile, the cross-correlation temporal profile of the coupling between the "original" electron pulse (~500 fs) only and the NIR laser pulse, in the absence of the "gating" visible pulse, is measured and the fitted curve is plotted in orange in Fig. 3d (see Supplementary information, Fig. S1).

From the difference of these temporal profiles, we obtained the cross-correlation profile of the coupling between the "gated" electron pulse and the NIR laser pulse. This temporal profile and its fitting are shown in (Fig. 3e) in connected solid black dots and solid red line, respectively. The FWHM duration of this profile is ~50 fs. Although this cross-correlation temporal profile is broader than expected between the "gated" electron pulse (~30 fs) and the NIR laser pulse (~33 fs) which should be on the order of ~45 fs, this broadening is likely due to slight saturation of the coupling between the NIR and "original" electron pulses (see Supplementary information, section I). However, since the "gating" optical pulse is kept below the saturation level, the "gated" electron pulse is likely close to 30 fs.

To demonstrate the full control of the "gated" electron pulse by controlling the optical-gating time window, we have repeated the same measurements under the same condition, but utilizing a longer visible "gating" laser pulse (~110 fs). The temporal profile of this pulse is measured and plotted as a blue line in Fig. 3f. The fitting of the retrieved cross-correlation temporal profile of the NIR laser pulse (~33 fs) and the "gated" electron pulse is shown by solid red line in Fig. 3f. The FWHM duration of this cross-correlation profile becomes ~175 fs, which indicates a longer "gated" electron pulse. The cross-correlation is longer than expected which is possibly due to the broadening of the cross-correlation dip which makes it more sensitive to the subtraction step.

It is noteworthy to mention again that these measurements have been conducted with low "gating" pulse intensity to reduce the saturation effect of the photon-electron coupling between the visible "gating" and "original" electron pulse, which could make the gating window longer and generate a longer "gated" electron pulse. Also, the gold nanoparticle was chosen because it has a very short surface plasmon lifetime[35] on the order of a few femtosecond so the intrinsic system dynamics will not interfere with our temporal characterization measurements.

The main objective of this work is to report a major enhancement of the temporal resolution (30 fs) for practical imaging in UEM. Certainly, the optical-gating technique provides such enhancement. Here, the generated "gated" electron pulse has sufficient electron counts (~8% of the total electron counts or < 1 electron/pulse) for probing ultrafast dynamics of matter. For example, Fig. 3 inset shows a plasmonic field image of a silver nanowire acquired in 30 seconds ($1.2 \times 10^5$ gated electrons) using the electron pulse gated by the 30 fs optical pulse. The attained temporal resolution (30 fs) allows resolving and imaging the electron dynamics with lifetimes on the order of a few tens of femtoseconds, for instance, the electron-electron scattering and the electron phonon coupling in semiconductors,[36] and the dynamics of surface plasmons which have long lifetimes (i.e. silver nanowire).[35] This can be done by utilizing another laser pulse to "pump" the system (nanostructure sample under study), together with the "gating" laser pulse to generate the ultrashort electron pulse which acts as a "probe" to image the system dynamics. The measurement can take place on the same stage (and same nanostructure sample) where the gating takes place to avoid any space charge effect on the "gated" electron pulse due to propagation.

**Attomicroscopy: Towards imaging the electron motion in real time**

As previously explained, the generated "gated" electron pulse duration is limited by the "gating" laser pulse duration. Hence, with the recent advance in ultrafast light science[37] and the generation of the optical attosecond pulse demonstrated in Hassan et. al,[11] the optical-gating approach utilizing this optical attosecond pulse promises the generation of subfemtosecond electron pulses. Accordingly, we have conducted a theoretical experiment (see Supplementary information, section II), where the demonstrated optical attosecond pulse[11] (Fig. 4a) is used to gate a pre- compressed electron pulse (75 fs). The multiphoton absorption probability has been calculated using the following equation[38].

$$P_L = \sum_{N=|L|}^{\infty} \sum_{N`=|L|}^{\infty} C_L^N \left(C_L^{N`}\right)^* \frac{\exp\left\{-\frac{(N+N`)(\tau/\tau_p)^2}{1+[(N+N`)/2](\tau_e/\tau_p)^2}\right\}}{\sqrt{1+\left[\frac{(N+N`)}{2}\right](\tau_e/\tau_p)^2}} \quad (1)$$

where N and N` are possible numbers of total scattering events, $C_L^N$'s are the expansion coefficients of the electron wave function in the basis of momentum eigenstates corresponding to gaining or losing a certain number of photons, $\tau$ the delay between the electron and "gating" optical pulses, $\tau_e$ the electron pulse duration, and $\tau_p$ the laser "gating" pulse duration.

A simulated electron energy spectrogram of the attosecond optical coupling is shown in Fig. 4b and the coupling electron energy spectrum at $\tau_{OAP} = 0$ fs is shown in Fig. 4c. The ZLP has been subtracted from both the spectrogram and the spectrum for better visualization of the coupling peaks. The gating time window emulates the attosecond optical "gating" pulse which permits the generation of isolated subfemtosecond electron pulses. For such a broadband "gating" optical pulse, the gating medium (nanostructure) should have a broad frequency response and the gated electron spectral peaks are expected to exhibit an energy broadening (Fig. 4 b &c). The ratio of the number of attosecond-gated electrons to the total number of electrons of the "original" electron pulse represents the attosecond optical gating efficiency which depends on the "original" electron pulse duration. It is calculated by dividing the integration of the coupling part of the spectrum by the integration of the entire spectrum. This relation is illustrated in Fig. 4d. Notably, these calculations have been conducted in low field intensity ($1.5 \times 10^{12}$ W/m$^2$) to avoid the temporal broadening of the "gated" electron pulse due to the saturation (see Supplementary information, Fig. S2). Thus, the attosecond optical gating of a short (few tens of femtoseconds) electron pulse will allow the generation of attosecond electron pulses with sufficient intensity to capture the electron motion in the act. Remarkably, the extreme temporal resolution stability for the potential pump-probe experiment utilizing the isolated attosecond "gated" electron pulse as a "probe" and a laser pulse as a "pump" can be achieved by the phase locking of the laser "pump" pulse and the optical "gating" pulse which is very common in the attosecond physics field. Also, the phase locking of the "gating" optical attosecond pulse is essential to minimize the timing jitter and the fluctuation of the "gated" electron pulse intensity. This advancement will open the way for attaining the attosecond resolution in electron microscopy and establishing the "Attomicroscopy".

The demonstrated optical-gating approach utilizing ~30 fs optical pulses enables more than an order of magnitude enhancement of the temporal resolution in electron microscopy, which permits the imaging of ultrafast electronic dynamics that last a few tens of femtoseconds, as for instance the electron dynamics in semiconductors and the metallic plasmon dynamics in nanostructures. Furthermore, this work demonstrates the potential of attosecond optical-gating to attain attosecond resolution in electron microscopy by generating isolated subfemtosecond electron pulses to image the electron motion in real time and space.

**Figures**

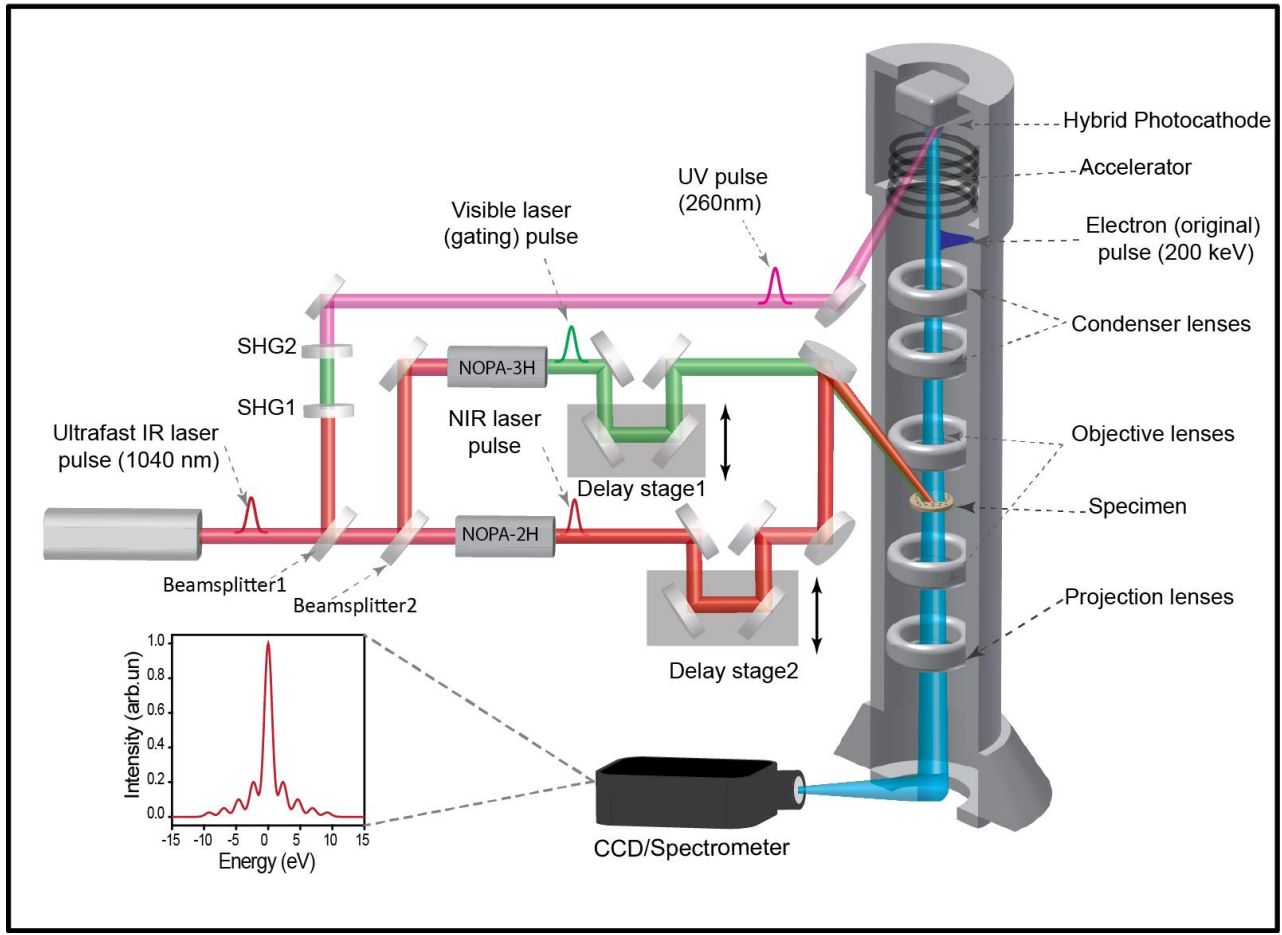

**Fig. 1. Experimental setup of the optical-gating in UEM**. A portion of infrared laser pulses generates ultraviolet (UV) laser pulses by two sequential second harmonic generation processes. These pulses are directed to the photoemissive cathode to generate ultrafast electron pulses. The second portion is divided into two equal beams sent to two NOPA systems to generate visible optical "gating" pulses (~30 fs) and near-infrared (NIR) laser pulses (~33 fs). The delays between these pulses are controlled by linear delay stages. These pulses are recombined and focused onto the specimen in the microscope and the electron energy spectrum of the electron pulses is acquired by electron energy spectrometer. An energy spectrum for electron-visible photon coupling is shown in the inset.

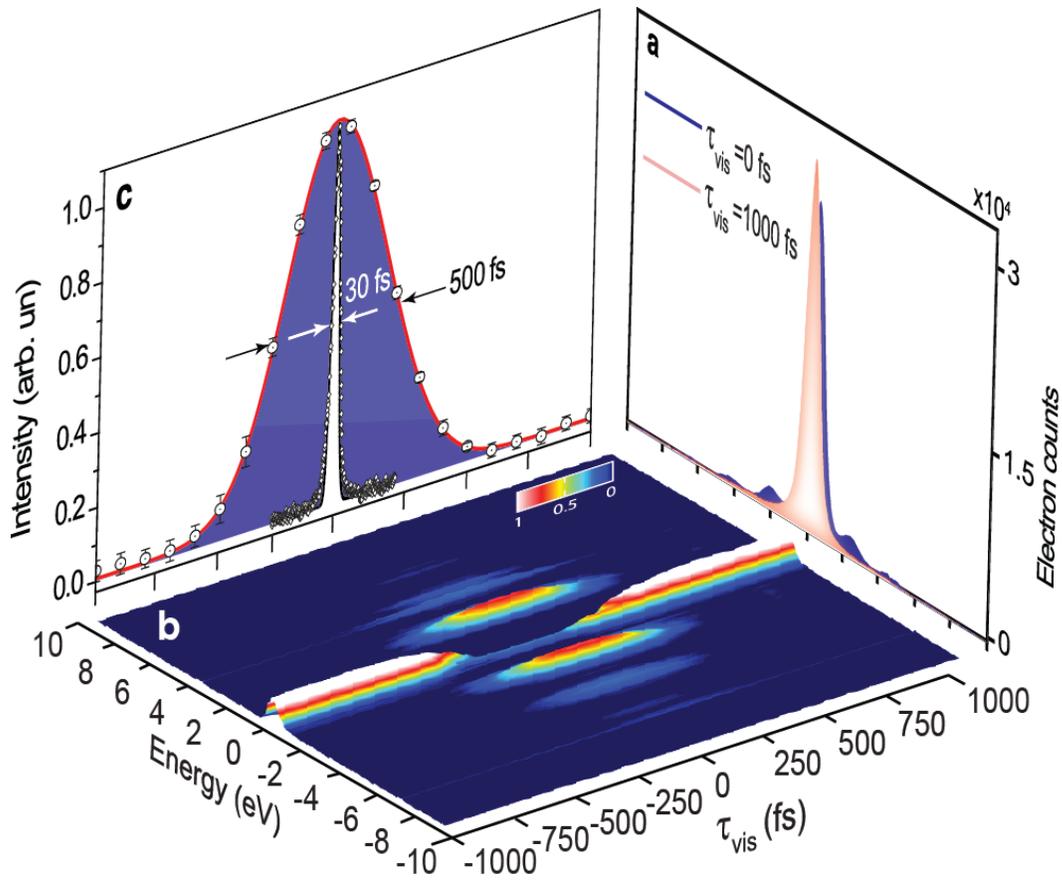

**Fig. 2. The temporal characterization of ultrafast electron pulse**. **a**, The electron energy spectrum of the "original" ultrafast electron pulse (ZLP spectrum) is shown in the shaded reddish white curve, and the electron energy spectrum with the coupling between the electron and "gating" visible (ℏω=2.25eV) laser pulse ($\tau_{Vis} = 0\ fs$) is plotted in blue. **b**, The measured electron energy spectrogram of the electron energy spectra as a function of the "gating" visible laser pulse delay $\tau_{Vis}$ where the gated electrons appear on both sides of the ZLP (energy gain and energy loss sides). The residual ZLP at $\tau_{Vis} = 0\ fs$ is subtracted for better visualization. **c**, The cross-correlation temporal profile, calculated by integrating the coupling spectral peaks of the energy spectrum at each instant of "gating" pulse arrival time ($\tau_{Vis}$), is shown in open black circles and its fitting is shown as a blue shaded red curve. The white shaded curve and white dotted line show the time of the gating window (the pulse duration of the "gating" visible pulse).

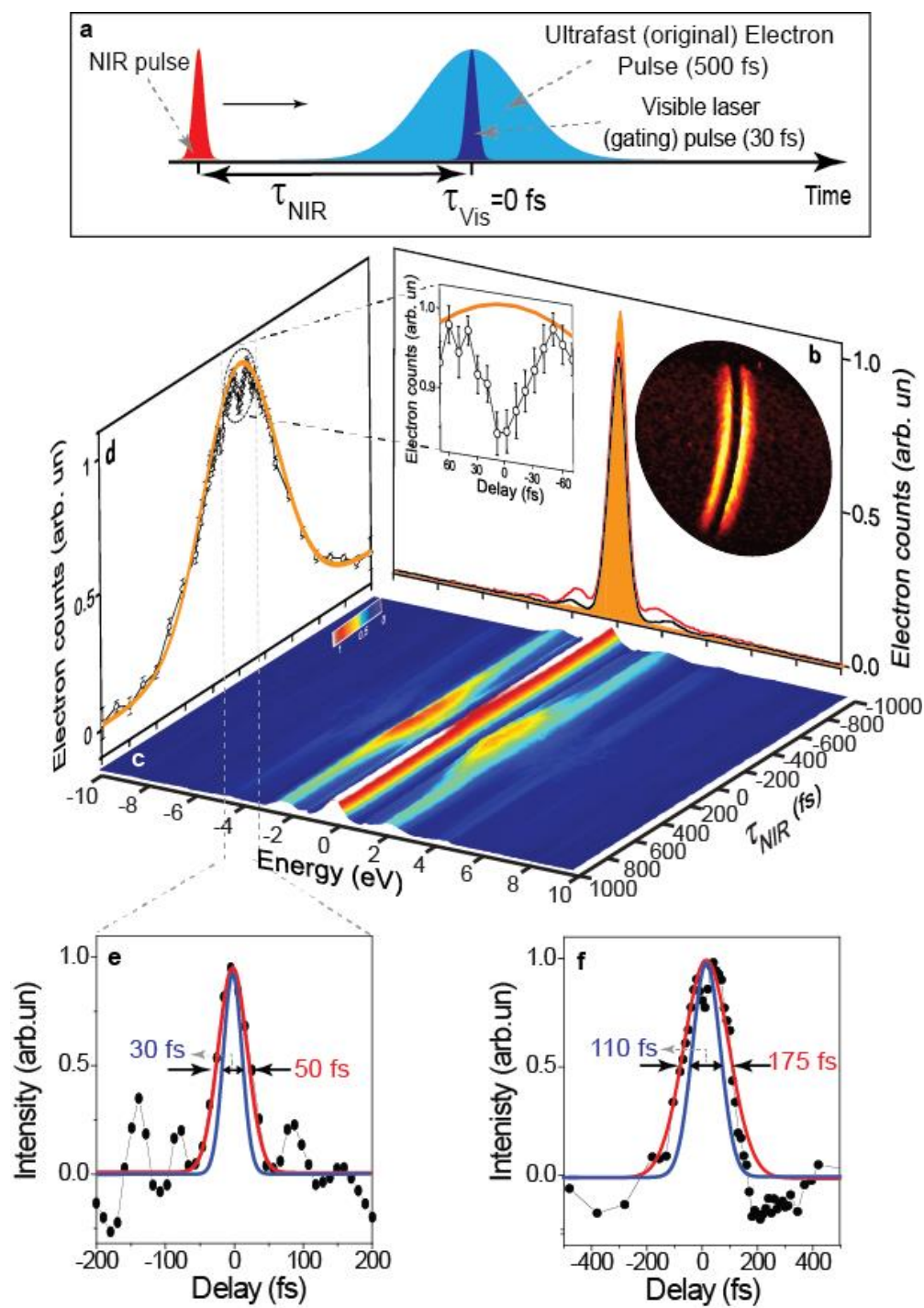

**Fig. 3. The temporal characterization of gated electron pulse**. **a**, Illustration of the principle of the cross-correlation measurement to characterize the temporal profile of the gated electron pulse. **b**, The electron energy spectra of optimum electron-photon coupling between "original" electron pulse and "gating" visible pulse (30 fs) (black line), the "original" electron pulse with both visible and NIR ($\hbar\omega=1.675$ eV) laser pulses (red line), and the "original" electron pulse spectra (ZLP) (orange curve). **c**, The cross-correlation electron energy spectrogram of the time dependence of electron-photon coupling between the NIR laser pulse and both "original" and "gated" electron pulses. The ZLP is suppressed for clear illustration of the gating effect. **d,** The cross-correlation temporal profile retrieved from the measured spectrogram in **(c)** carries the signature of the "gated" electron pulse through the coupling of the "original" electron pulse with NIR laser pulse, shown by connected open circles. This curve and its expanded view (inset in **(b)**) show clearly the dip in the electron counts due to the "gated" electron pulse. The orange line curve represents the fitting of the measured cross-correlation temporal profile of NIR pulse coupling with the "original" electron pulse in the absence of "gating" visible pulse. **e**, The cross-correlation of the "gated" electron and NIR pulses, obtained by subtraction of the temporal profiles in (**d**), is plotted in black dots along with its fitting (red line). A fit of a measured temporal profile of the "gating" visible pulse is shown in blue. **f**, Same as (**e**) for the case of a longer visible "gating" pulse (~110 fs).

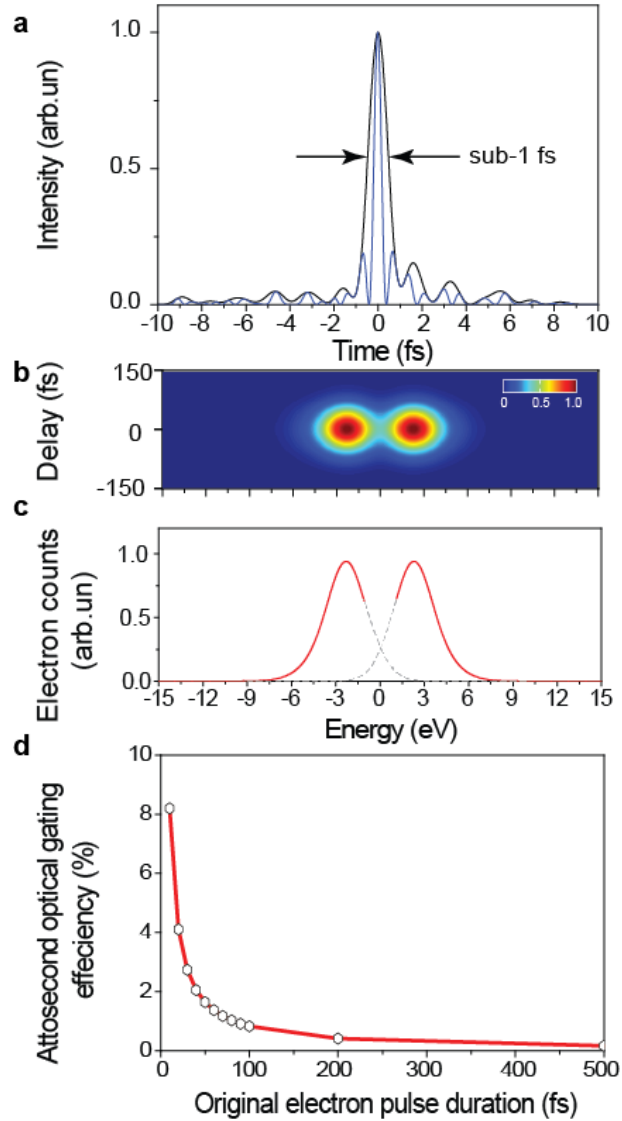

**Fig. 4. Attosecond optical gating of the electron pulse**. **a**, The optical attosecond "gating" pulse. **b**, The simulated optical attosecond gating spectrogram of a 75 fs electron pulse by the optical attosecond pulse at intensity below the saturation level. **c**, The electron energy spectrum at $\tau_{OAP} = 0\ fs$. Note, the ZLP in **b** and **c** is subtracted for better visualization. **d**, The attosecond optical gating efficiency as a function of the pulse duration of the "original" electron pulse.

**Method**

In the optical-gating experimental setup, illustrated in Fig. 1, a train of IR (350 fs-120 µJ) laser pulses (λ ~1040 nm, 100 kHz repetition rate) is divided into three beams, with 5% going through a second harmonic process twice to generate ultraviolet (UV) laser pulses (λ ~260 nm). These UV pulses are directed to the photoemissive cathode inside the microscope to generate ultrafast electron pulses ("original" electron probe pulse), which are accelerated (200 keV) in the microscope column. The rest of the IR (95%) is divided by a 50/50 beamsplitter into two beams. Each beam enters into a different non-collinear optical parametric amplifier (NOPA) system. In the first one (NOPA-3H), the visible laser pulses are generated (λ ~550 nm). These pulses are compressed and have a FWHM pulse duration of ~30 fs. The second one (NOPA-2H) generates ~33 fs laser pulses in the near-IR (NIR) region (λ ~740 nm). The delays between the electron, visible, and NIR pulses are controlled by linear delay stages. The visible laser pulse acts as a "gating pulse" to generate a "gated" electron pulse emulating its pulse duration (~30 fs), and the NIR pulse is utilized to characterize the temporal profile of the gated electron pulse. Notably, to ensure the short pulse durations of visible and the NIR pulses (~30, ~33 fs, respectively) at the sample position inside the microscope, the two laser pulses are precompensated by a built-in prism compressor implemented inside each NOPA system for the dispersion accumulated due to the laser beams traveling through the air and glass up to the sample position. The correct compression was determined by maintaining an equivalent amount of glass in the paths to the microscope (beam splitters, lens, and microscope window) and to the autocorrelator apparatus used to characterize the optical pulses' temporal profiles.

**Acknowledgment**

We would like to thank G. M. Vanacore and T. Karam for fruitful discussion. This work was supported by the National Science Foundation Grant DMR-0964886 and the Air Force Office of Scientific Research Grant FA9550-11-1-0055 for research conducted in The Gordon and Betty Moore Center for Physical Biology at the California Institute of Technology.

**Materials& Correspondence**

Corresponding Authors: M. Th. Hassan, email: mhassan@caltech.edu and J. S. Baskin, email: baskin@caltech.edu

**Author Contributions**

M. Th. H. conceived the idea; M. Th. H., J. S. B. and A. H. Z designed the experiment; M.Th. H., and J. S. B. conducted the experiments; M. Th. H., J. S. B. and A. H. Z. conducted the analysis of the first set of results. M.Th. H., and B.L. conducted the simulations; B.L and J. S. B. and M. Th. H. interpreted the data and contributed to the preparation of the manuscript.